\documentclass[12pt]{iopart}

\usepackage{epsfig}

\begin{document}

\title{SQM 2006: Theory Summary and Perspectives}
\author{Steffen A.~Bass\dag}
\address{\dag\ Department of Physics, Duke University, 
             Durham, North Carolina 27708-0305, USA}

\ead{bass@phy.duke.edu}

\submitto{\JPG}



\section{Introduction}

The investigation of strangeness production in relativistic heavy 
ion collisions has been proven to be a powerful tool for the study 
of highly excited nuclear matter, both in terms of the reaction
dynamics and in terms of its hadrochemistry
\cite{Rafelski:pu,koch86,koch88,Braun-Munzinger:1994xr,Letessier:ad,Vance:1999pr,Soff:1999et,Braun-Munzinger:1999qy,Rafelski:2001hp,Braun-Munzinger:2001ip}.
Furthermore, strangeness has  been suggested as a signature
for the creation of a Quark-Gluon-Plasma (QGP) 
\cite{Rafelski:pu,koch86,koch88}:
in the final state of a heavy-ion collision, subsequent to the formation
and decay of a QGP, strangeness has been predicted to
be enhanced relative to the strangeness yield in
elementary hadron+hadron collisions. It was actually this early work
on strangeness as a QGP signature which was instrumental in creating
the SQM conference series. However, as witnessed by the work presented
at this conference and its predecessors, strangeness production
in heavy-ion collisions has developed into an
impressively versatile tool for the characterization of QCD matter.

In this write-up I will discuss a selection of key contributions at
SQM 2006 which I consider to have a large impact to the current
status of the field of strangeness physics 
or which may have the potential to significantly
advance strangeness -- or in general flavor physics -- 
in the near future.

I would like to point out that this write-up is not a comprehensive summary covering
all theory topics discussed at SQM 2006 -- for example, there have been excellent
mini-reviews on strangeness in the nuclear astrophysics context by 
J. Schaffner-Bielich \cite{Schaffner} and S. Reddy \cite{Reddy} as well as on
the extraction of the nuclear equation of state via near-threshold
kaon production by J. Aichelin \cite{Aichelin} and the extension of the periodic table
into the strangeness and anti-matter sector by W. Greiner \cite{WGreiner} -- the
coverage of any of these would be beyond the scope of this write-up.

\section{State of the Art: Lessons and Challenges}

\subsection{Hadron Yields and Ratios}

The impressive success of 
statistical models in describing the (strange) hadron
abundances and ratios at the CERN/SPS and RHIC
\cite{Braun-Munzinger:1994xr,Letessier:ad,Braun-Munzinger:1999qy,Rafelski:2001hp,Braun-Munzinger:2001ip}.
and the extracted
$\gamma_s$ values close to 1
have led to the common conclusion that chemical freeze-out in 
ultra-relativistic heavy-ion reactions
 occurred very close to -- or even at hadrochemical equilibrium
and that this state most likely has been created by a hadronizing QGP. 
Several
studies, exploring the systematics of hadron ratios as function
of collision system and energy have been presented at this meeting, all
corroborating the model's success and previous findings
\cite{sqm06_SM}.

However, there remain a number of open questions associated with
the success of the statistical model, many of which transcend the
scope of the model and need to be addressed either through direct
analysis of data beyond yields and spectra or via 
dynamical non-equilibrium 
theory approaches based on ab-initio calculations:
\begin{itemize}
\item how did the system achieve chemical equilibrium?
{\em Conventional }calculations based on boost-invariant hydrodynamics
with rate-equations for quark production
\cite{kapusta86a,matsui86a,elliott00a},
pQCD rate-equations \cite{biro93a} or
the Parton Cascade Model \cite{geiger93a} all indicate that chemical
equilibration (and
strangeness saturation) cannot be achieved during realistic life-times
of the deconfined phase. It has been suggested (e.g. in
\cite{kapusta86a,matsui86a}) that the system would be driven toward and
come close to chemical equilibrium in the subsequent hadronic phase
-- a scenario which would help to bridge the gap between the calculations
indicating insufficient equilibration time in the plasma phase and the
copious SPS and RHIC
data apparently close to chemical equilibrium at chemical freeze-out.
Recent calculations assuming a hadronizing QGP out of chemical equilibrium
with subsequent hadronic rescatting have shown that rescatting via binary
collisions in the hadronic phase is insufficient to drive the system toward
chemical equilibrium before the expansion of the system leads to chemical
freeze-out \cite{bass_sqm00}.
Currently the most favored approaches for explaining the rapid thermalization
of the medium are either centered around turbulent color fields
\cite{Mrowczynski:1988dz,Mrowczynski:1993qm,Romatschke:2003ms,Randrup:2003cw,Arnold:2003rq} 
or multi-particle collisions (which may take place either in the deconfined
or dense, confined phase of the reaction) \cite{Xu:2004mz,rapp,greiner,Braun-Munzinger:2003zz}. 
A smoking gun signature for either mechanism remains yet to be established.
\item do the temperature
and chemical potential extracted from the
statistical model fits to final state hadron yields or ratios really reflect
the thermodynamic state of the system at one particular time during its
evolution (i.e. the conditions at chemical freeze-out) or are they rather
the result of a superposition of different states, due to individual
hadron species decoupling continuously from the system (as would
be expected from their different mean free paths)? The latter
view is supported by a transport model analysis of the time-evolution
of the temperature and chemical potential in the central
cell of a heavy-ion reaction \cite{bravina}. However, one would
expect a fairly large chi-squared for the single source statistical
model fit in case of a continuous decoupling scenario, which is not observed
in the data. It should also be noted that the use of $4\pi$ integrated 
yields at SPS energies creates ambiguities in the fit, since many hadron
ratios have been shown to exhibit a strong rapidity dependence at 
SPS (and lower) beam energies \cite{Bass:1997xw,Sollfrank:1998zg}.
True progress on this issue can only be
achieved if a model-independent method is found to determine the
chemical decoupling time of individual hadron species.
\item taking the extracted temperature and chemical potential values
for SPS and RHIC at face value, the system chemically decouples
very close to $T_C$ -- at a temperature at which the properties
of hadrons (e.g. their mass and/or width) 
could still be substantially modified by the temperature and density
of the medium (see e.g. the spectral function of the $\rho$ presented
at this meeting by A. Foerster \cite{Foerster}). 
Is the use of hadronic vacuum masses and
widths in the statistical model the right approach or should their
temperature- and density dependent medium modifications be taken
into account in the model fits \cite{Michalec:2001qf,Zschiesche:2002zr}?
\end{itemize}

Overall, to this day the physics mechanisms and driving forces behind
the impressive  success of the statistical model are not well understood
-- finding an answer to the question
{\em why} the statistical model performs so well is a challenge the
theory community must address with renewed vigor in the near future.

\subsection{Excitation Functions}

Over the past decade the experimental programs at the AGS, SPS and RHIC have
resulted in a wealth of data which can be compiled into excitation
functions of yields, ratios, flow coefficients, system sizes etc.. However,
the hope of finding some kind of sharp discontinuity as a signal of a 
phase transition has not come to fruition. This may be partially due to
many QGP signatures predicted to exhibit a sharp discontinuity or local extrememum
relying heavily on the deconfinement phase-transition being of first order
with a long-lived mixed phase -- an assumption which is at odds with recent
findings from Lattice QCD, predicting the phase transition to be a continuous
crossover
in the RHIC and SPS domain. Even in the case of a 1st order phase-transition
the expectation of sharp features in excitation functions may have been
somewhat naive: sharp discontinuities are predicted as function of temperature, which
does not have a one-to-one relation to beam energy.
Realistic temperature and density profiles as well as 
corona effects (see e.g. the contribution by K. Werner \cite{Werner}) 
would lead to a smearing out of
the features associated with deconfinement and thus would make for a
smooth variation in the excitation function.

Spurred by recent lattice-gauge theory calculations of QCD at finite
baryon density \cite{Gavai} there has been renewed hope that an excitation
function in incident-beam energy might yield interesting results if one
can create a QCD medium close to the tri-critical point.  It has been
found that the net-quark susceptibility near the tri-critcial point diverges,
which would be experimentally accessible via charge fluctuation measurements 
\cite{Ejiri:2005wq,Redlich}. 

\subsection{Transport Theory}
Relativistic Fluid Dynamics (RFD, see e.g.
\cite{Bjorken:1982qr,Clare:1986qj,Dumitru:1998es})
is ideally suited for the high-density phase of heavy-ion reactions
at RHIC,
but breaks down in the later, dilute, stages of the
reaction when the mean free paths of the hadrons become
large and flavor degrees of freedom
are important. The biggest advantage  of  RFD
is that it directly incorporates an equation of state as input
and thus is so far the only dynamical model in which a phase
transition can
explicitly be incorporated. In the ideal fluid approximation
(i.e. neglecting off-equilibrium effects) and once the initial conditions
for the calculation have been fixed,
the EoS is the {\em only}
input to the equations of motion and relates directly to
properties
of the matter under consideration. Ideally, either the initial conditions or the 
EoS should be determined beforehand by an ab-initio calculation (e.g. for the EoS via
a lattice-gauge calculation), in which case a fit to the data would allow
for the determination of the remaining quantity.

RFD has been extremely successful in describing single
particle spectra and collective flow effects at RHIC
\cite{Kolb:2003dz,Huovinen:2003fa,Hirano:2002hv}, even though no
hydrodynamical model implementation has so far attempted to address the
entire array of available data in a single consistent calculation
(a forthcoming publication aims to remedy this situation \cite{Nonaka:2006}).
The shape of the spectra as well as the
transverse momentum dependence of the elliptic flow for
minimum bias data are generally reproduced nicely. However, more
specific centrality bins pose a problem for the elliptic
flow calculations. One should also bear in mind that
hydrodynamical calculations which assume 
a standard chemical equilibrium hadron gas equation of state 
below $T_c$ (which implies simultaneous chemical and 
thermal freeze-out) are unable to fit
the measured particle yields simultaneously together
with the spectra, since statistical models show chemical
freeze-out to occur around T=170~MeV, whereas the shape
of the spectra requires a hydrodynamic evolution to
T=110~MeV. 
One method to deal with the separation of chemical and thermal freeze-out
is the {\em partial chemical equilibrium} model (PCE) \cite{Arbex:2001vx,Hirano:2002ds,Kolb:2002ve}.
Below a chemical freeze-out temperature $T_{\rm ch}$ one introduces a chemical
potential for each hadron whose yield is supposed to be frozen out at $T_{\rm ch}$.
The PCE approach can account for the proper normalization of the spectra,
however, it fails to reproduce the transverse momentum and mass dependence
of the elliptic flow.
More importantly for the
strangeness sector, ideal RFD calculations lack the capability
of dealing with the flavor dependence of hadronic cross sections
and a possible flavor-dependent sequential freeze-out:
recent experimental results suggest that
at thermal freeze-out multistrange baryons exhibit less transverse flow
and a higher temperature closer to the chemical freeze-out
temperature compared to non- or single-strange baryons
\cite{STAR_strange1,STAR_strange2}. This behavior can be understood in terms
of the flavor dependence of the hadronic cross section, which decreases
with increasing strangeness content of the hadron. The reduced
cross section of multi-strange baryons leads to a decoupling from the
hadronic medium at an earlier stage of the reaction, allowing them
to provide information on the properties of the hadronizing QGP less
distorted by hadronic final state 
interactions \cite{vanHecke:1998yu,Dumitru:1999sf}.

The reach of RFD can be extended and the problem
having to terminate the calculation at a single flavor-independent
and fixed freeze-out temperature 
can be overcome by combining the RFD calculation
with a microscopic hadronic cascade model -- this kind of
hybrid approach (dubbed {\em hydro plus micro}) was pioneered
in \cite{Bass:2000ib} and has subsequently been adopted by other
groups \cite{Teaney:2000cw,Hirano:2005vw,Nonaka:2005aj,Nonaka:2006}.
Its key advantages are that the freeze-out occurs
naturally as a result of the microscopic evolution and that
flavor degrees of freedom are treated explicitly through the
hadronic cross sections of the microscopic transport.
Due to the Boltzmann equation being
the basis of the microscopic calculation in the hadronic phase,
viscous corrections for the hadronic phase are by default
included in the approach -- the full treatment of viscosity
in the deconfined phase in a
3D hydrodynamic calculation remains a challenge for the future.

Such hybrid macro/micro transport calculations  are to date the
most successful approaches for describing the soft physics
at RHIC \cite{Hirano:2005wx}. First implementations of these hybrid approaches
were restricted to 1+1 \cite{Bass:2000ib} and
2+1 \cite{Teaney:2000cw} dimensions
in the hydrodynamic component of the model.
However, recently new state of the art  fully three-dimensional ideal
hydrodynamics models have become available and are now being
incorporated into the hydro+micro framework \cite{Nonaka:2005aj,Hirano:2005xf,Hirano:2005vw,Nonaka:2006}.

The latest model implementation has been shown at this conference \cite{Nonaka:2006} to be 
able to excellently reproduce the measured $P_T$ spectra 
and $\langle P_T \rangle$ values of hyperons and multi-strange
baryons at RHIC (see left frame of figure~1). A collision
number analysis for these hadron species based on
this calculation \cite{Nonaka:2006} 
confirms the early and sequential freeze-out findings
described in \cite{vanHecke:1998yu,Dumitru:1999sf}.

\begin{figure}[t]
\centerline{
\psfig{file=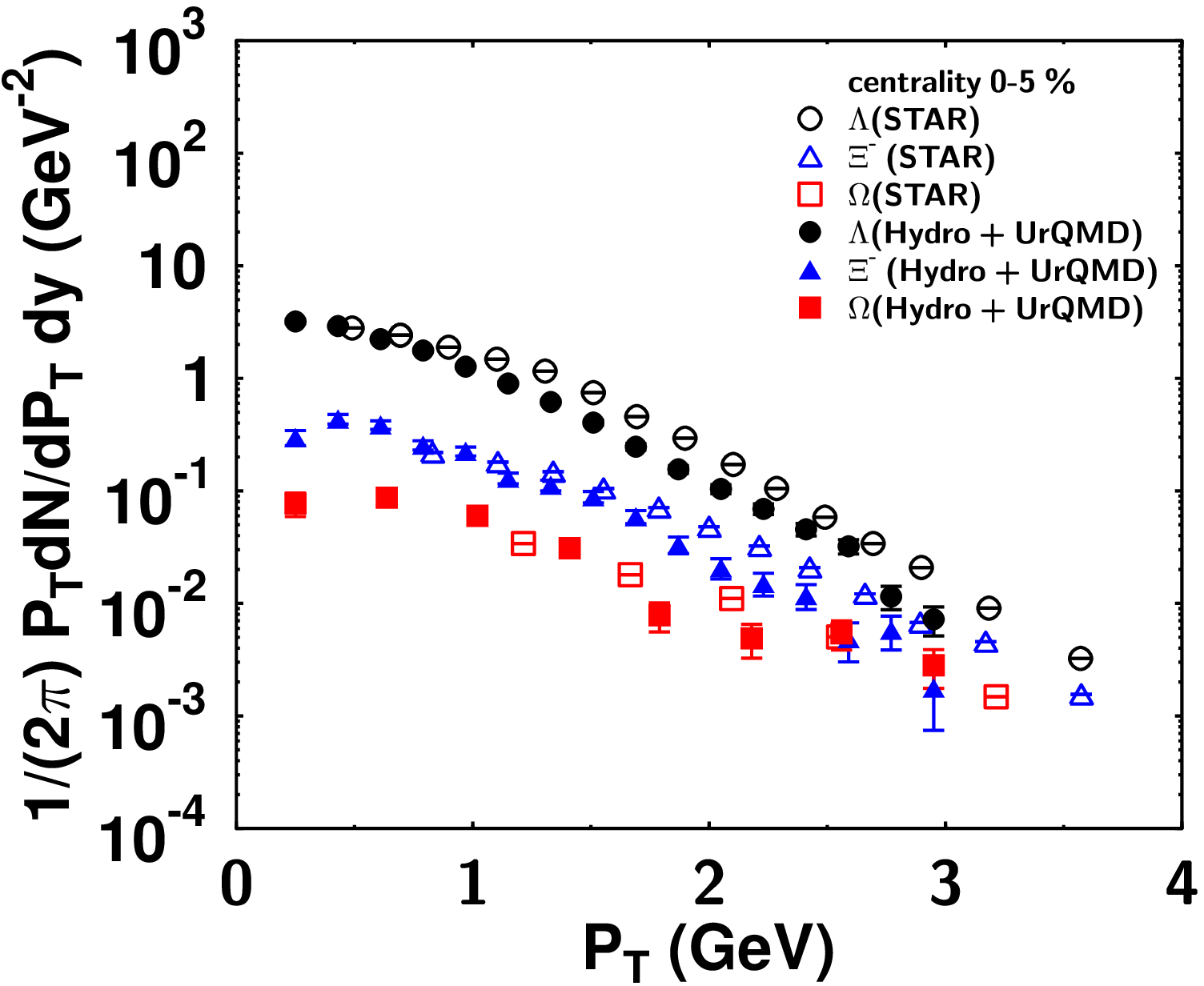,width=8cm}
\psfig{file=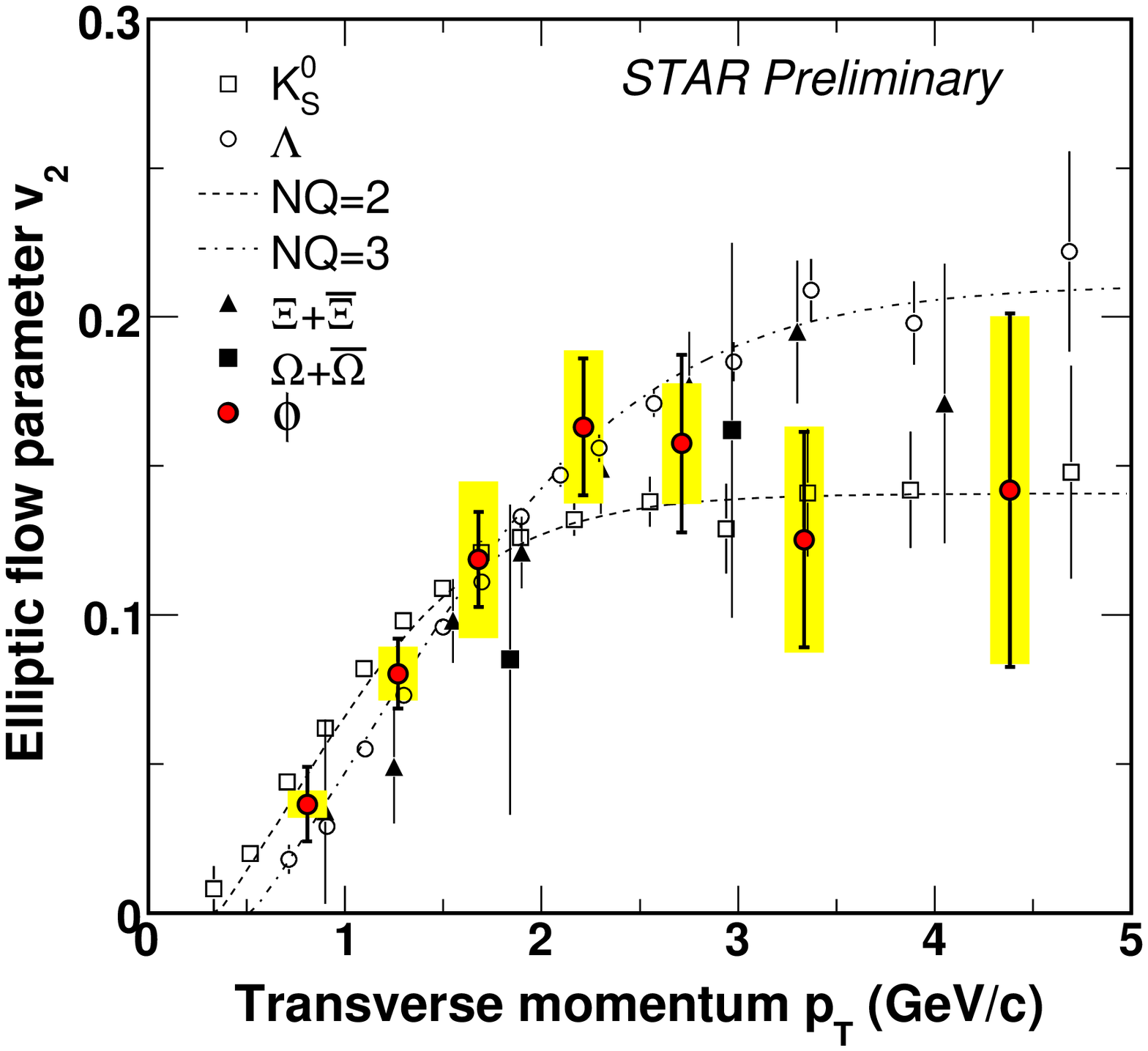,width=8cm}
}
\vspace*{8pt}
\label{hu_pt}
\caption{Left:
$P_T$ spectra for $\Lambda$, $\Xi$, $\Omega$ in central collisions
calculated in the hybrid 3D-hydro+UrQMD approach
compared to STAR data. Right: elliptic flow of (multi-)strange hadrons
measured by the STAR collaboration \cite{slblyth,Adams:2005zg}. Note
the $\phi$ meson following the kaon elliptic flow. }
\end{figure}

Marcus Bleicher discussed elliptic flow at RHIC in the framework of 
the microscopic hadron/string 
UrQMD and RQMD transport models \cite{Lu:2006qn,bleicher}.
Both models reach only 60\% of the absolute magnitude 
of the measured $v_2$, due to the string degrees of freedom
being used to describe the initial phase in these models 
generating insufficient pressure. Interestingly, the model
calculations exhibit features reminiscent of constituent
quark number scaling in the intermediate $p_T$ range between 2 and 6 GeV/c. 
This is due
to the use of the Additive Quark Model for most hadron-hadron
cross section implemented in the model, thus having the
interaction scale with the number of constituent quarks.
At closer inspection, however, one finds a flavor ordering
in the elliptic flow generated by the microscopic 
transport models: $v_2(N) > v_2((Y) > v_2(\Xi) > v_2(\Omega)$,
due to the strange quark being assigned a smaller interaction
cross section than the $u$ and $d$ quarks. This structure
is not observed in the data, which rather indicate the
build-up of elliptic flow independent of the quark flavor
prior to hadronization.

\subsection{Hadronization via Parton Recombination}

The detailed understanding of hadronization plays a crucial role
for isolating signatures sensitive to the QGP evolution and properties
of the system from those which are dominated by the later
reaction stages. One of the theoretical milestones of the first
several years of the RHIC program was the development of the
recombination plus fragmentation model as the standard model
of hadronization for matter at intermediate and high transverse
momenta: at the center of the recombination + fragmentation model is the
realization that hadron production at momenta of a few GeV/$c$ in an
environment with a high density of partons occurs by recombination,
rather than fragmentation, of partons. It is found that recombination
always dominates over fragmentation for an exponentially falling parton
spectrum, but that fragmentation wins out eventually, when the spectrum
takes the form of a power law.

Among the RHIC discoveries that prompted the development of the 
recombination models of hadronization,
is that the  amount of suppression of hadron
production at intermediate $p_T$ compared to the scaled
proton-proton baseline seems to 
depend on the hadron species.
In fact, in the production of protons and antiprotons between 2 and
4 GeV/$c$ the suppression seems to be completely absent. Generally,
pions and kaons appear to suffer from a strong energy loss while
baryons and antibaryons do not. Two stunning experimental facts
exemplify this \cite{Adcox:2001mf,Adler:2003kg}. First, the ratio of
protons over positively charged pions is equal or above one for $p_T
> 1.5 \rm{ GeV}/c$ and is approximately constant up to 4 GeV/$c$.
Second, the nuclear suppression factor $R_{AA}$ below 4 GeV/$c$ is
close to one for protons and lambdas ~\cite{Adams:2003am}, while it
is about 0.3 for pions.

The recombination
approach~\cite{FMNB:03,Fries:2003kq,GreKoLe:03,HwaYa:02} has been
able to account for the above baryon/meson differences.
Additionally, it was observed that the elliptic flow pattern of
different hadron species can be explained by a simple recombination
mechanism \cite{LinKo:02,Voloshin:02,MoVo:03,LinMol:03}. The
anisotropies ,$v_2$, for the different hadrons are compatible with a
universal function $v_2(p_T)$ in the parton phase, related to the
hadronic flow by factors of two and three depending on the number of
valence quarks \cite{Sorensen:03,Adler:2003kt}.

One of the perceived weaknesses of the recombination+fragmentation
approach was that its development was triggered by the experimental
observation of the meson/baryon anomalies in the RHIC data and therefore
its success in explaining these features occurred after the fact.
However, the recombination approach was able to {\em predict} the 
the elliptic flow $v_2$ of multi-strange hadrons such as the
$\phi$, $\Xi$ and $\Omega$ as a function of transverse 
momentum \cite{Nonaka:2003hx}:
the measurement of $v_2$ for the $\phi$ and $\Omega$
allows for the  unambiguous distinction
between parton recombination and statistical
hadro-chemistry to be the dominant process in hadronization
at intermediate transverse momenta, since e.g. the $\phi$ meson
has approximately the mass of a nucleon, but the valence-quark content
of a meson. In a hydrodynamic picture with the hadron mass as the
guiding scale the $\phi$ would follow the systematics of the nucleon
whereas in the recombination picture it would follow the
behavior of the pions and kaons.
Data on the elliptic flow of the $\phi$ meson and $\Omega$ baryon
presented at this conference, e.g. in the talk by S.L. Blyth -- see
the right frame of figure~1 -- 
clearly exhibit constituent quark scaling
and therefore impressively confirm the physics of the
recombination model and demonstrate its predictive power.

\section{New Directions}

\subsection{Quarks in the Color Glass}

Heavy-ion collisions at ultra-relativistic
energies can be described by the collision of two coherent sheets
of high energy density gluonic fields (commonly referred to as 
{\em Color Glass Condensate}) \cite{McLerran:1993ka,McLerran:1993ni,Iancu:2003xm}.
Since the physical density of gluons becomes large, their typical 
separation is small, implying a small value for $\alpha_s$.
Furthermore, these highly coherent gluons saturate the phase space up to 
the maximal occupation number $\sim 1/\alpha_s$.
Due to the weak coupling, it is possible to
describe this system from first principles in QCD.

The {\em Color Glass Condensate} (CGC) has been suggested to describe the 
initial state of gold nuclei in RHIC collisions. While there
is a broad consensus that at sufficiently high beam energies
the saturation physics of the CGC should dominate the initial
state, it is still a matter of debate whether these conditions 
are actually fulfilled at RHIC. Nonetheless, the further 
development of the CGC picture for heavy-ion collisions is of 
great theoretical importance. Original work on the CGC
focused on the description of the initial state as gluonic field.
Due to the large gluon densities at low Bjorken-x, these are 
thought to dominate the dynamics of the initial state. However,
the lack of treatment and consideration of quark and anti-quark
production makes it difficult to connect the CGC to experimental 
data, e.g. in the strangeness sector.

T. Lappi presented a numerical integration of the Dirac equation in
order to calculate the
number of quark-antiquark pairs initially produced in the 
classical color fields of colliding ultra-relativistic nuclei (the
resulting state of high energy and density matter being termed
the {\em glasma}) \cite{Gelis:2005pb}. 
While the number of $q \bar q$ pairs is parametrically suppressed in 
the coupling constant, he found that in the CGC
their production rate is comparable to the thermal 
ratio of gluons/pairs $= 9 N_f/32$. After isotropization one 
thus would end up with a quark-gluon plasma in chemical equilibrium.
This finding is of great significance since it provides a link
between the CGC initial state and an abundance of flavor-centric
data indicating the formation of a QGP in thermal and chemical
equilibrium.

\subsection{Conserved Charge Correlations}
The elementary electric charge, strangeness and baryon number 
carried by the QGP degrees
of freedom  differs significantly from those in a gas of hadrons. This
observation does not only apply to the elementary charges themselves,
but also to the size of the average fluctuations of net baryon number, 
strangeness and electric charge in a finite volume.
While hadronization and confinement prohibit us from directly observing the
fractional electric charge and baryon number and strangeness 
of the QGP degrees of freedom,
these event by event fluctuations may under certain conditions
survive hadronization and subsequent hadronic rescattering and thus serve
as indicators of the existence of a QGP \cite{Asakawa:2000wh,Jeon:2000wg}.

Surprisingly, all experimental analysis of net charge fluctuations for
SPS and RHIC data agree with the hadron gas prediction, giving no
indication at all about a possible deconfined phase. Since many other
measurements are compatible with the assumption of deconfinement, it
is tempting to speculate that the dynamics of hadronization and/or
hadronic final state interactions strongly affect the charge fluctuation
observable and mask the fluctuations generated in the deconfined phase.

A more robust approach of exploiting the difference in conserved elementary
charges between QGP and HG degrees of freedom lies in the area of correlations:
in particular, the correlation between baryon number $B$ and strangeness $S$
may elucidate the microscopic structure of the QGP and the nature of its
degrees of freedom \cite{Koch:2005vg,Majumder:2006nq,Majumder}.
Strangeness is carried exclusively by $s$ and $\bar{s}$ quarks
which carry baryon number as well: $B_s= -\frac{1}{3}S$, whereas in a hadron
gas the relation between $B$ and $S$ is much less intimate.
This $C_{BS}$ correlation can be directly extracted from Lattice
QCD calculations and is experimentally accessible via event-by-event
fluctuations:
\begin{equation}
\label{CBS_eq1}
C_{BS}\equiv -3\frac{\sigma_{BS}}{\sigma^2_S} = -3 \frac{\langle BS\rangle
- \langle B \rangle \langle S\rangle}{\langle S^2 \rangle - \langle S\rangle^2}
= -3 \frac{\langle BS \rangle}{\langle S^2\rangle}
\end{equation}
For a system of hadrons $C_{BS}$ can be formulated in terms of multiplicity
variances $\sigma^2_k \equiv \langle n^2_k\rangle - \langle n_k \rangle^2 \approx \langle n_k \rangle$ for the respective hadron species $k$:
\begin{equation}
C_{BS} = -3 \frac{\sum\limits_k \sigma^2_k B_k S_k}{\sum\limits_k \sigma^2_k S^2_k}
\approx -3 \frac{\sum\limits_k \langle n_k \rangle B_k S_k}{\sum\limits_k \langle n_k \rangle S^2_k}
\end{equation}
For an ideal gas of hadrons at $T=170$~MeV and zero chemical 
potential $\mu_B=0$ one finds $C_{BS} = 0.66$.

The $C_{BS}$ correlator can also be expressed in terms of susceptibilities,
$C_{BS} = -3 \chi_{BS}/\chi_{SS}$ which are second derivatives of the free
energy with respect to the chemical potential and can be directly be
calculated from lattice QCD:
\begin{equation}
\chi_{BS} = -\frac{1}{V}\,\frac{\partial^2 F}{\partial \mu_B \partial \mu_S}\,,
\qquad \chi_{SS}=-\frac{1}{V}\,\frac{\partial^2 F}{\partial \mu_S^2}
\end{equation}
Using values of lattice susceptibilities extracted at $T=1.5\, T_C$ in
\cite{Gavai:2005yk} on finds $C_{BS}\approx 1$ indicating that the degrees of freedom
of a QGP carry baryon number and strangeness of individual quarks and suggesting that the quark flavors are uncorrelated as in an ideal QGP (see left frame
of figure~2).  Note that the presence
of pure gluon clusters cannot be ruled out by this diagnostic.
This finding is of great importance, since the evaluation of $C_{BS}$ for different
models of the QGP structure allows for the direct verification or falsification
of these models when compared to data and lattice QCD.

\begin{figure}[t]
\centerline{
\psfig{file=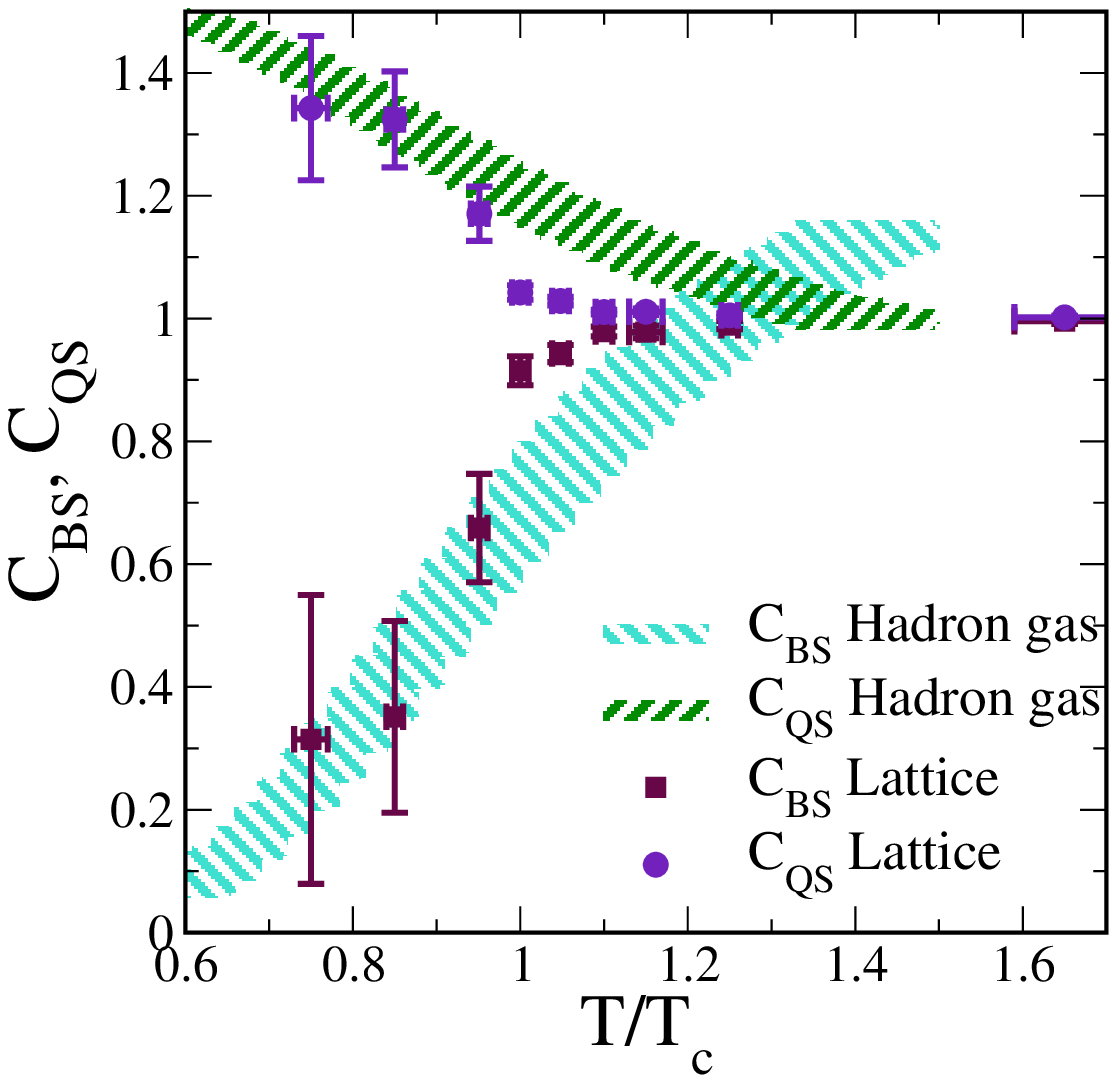,width=8cm}
\psfig{file=v2_e_MB_therm-av-qm05data.eps,width=8cm}
}
\vspace*{8pt}
\label{cbm_charm_v2}
\caption{Left: $C_{BS}$ and $C_{QS}$ in a truncated hadron gas at
$\mu_B=\mu_S=0$~MeV compared to lattice calculations at $\mu=0$.
The two bands denoting the hadron gas calculation and reflect the
uncertainty in the actual value of the phase transition temperature
$T_C$, which is assumed to lie in the range of $170\pm 10$~MeV 
(figure taken from \cite{Majumder:2006nq}).
Right: elliptic flow $v_2$ of non-photonic electrons in central Au+Au
collisions at RHIC. PHENIX data are compared to theory predictions
using Langevin simulations with elastic $c$ and $b$ quark interactions
in an expanding QGP (figure taken from \cite{Rapp:2006ta}).
 }
\end{figure}

\subsection{Anomalous Viscosity}

As stated earlier, measurements of the anisotropic collective flow of hadrons emitted
in non-central collisions of heavy nuclei at the Relativistic Heavy
Ion Collider (RHIC) are in remarkably good agreement with the
predictions of ideal relativistic fluid dynamics.
In order to describe the data, calculations need to assume that the
matter formed in the nuclear collision reaches thermal equilibrium
within a time $\tau_{\rm i} < 1$ fm/c \cite{Heinz:2001xi} and then
expands with a very small shear viscosity $\eta \ll s$, where $s$ is
the entropy density \cite{Teaney:2003pb}. The comparison between
data and calculations indicates that the viscosity of the matter
cannot be much larger than the postulated lower bound
$\eta_{\rm min} = s/4\pi$ \cite{Kovtun:2004de}, which is reached in
certain strongly coupled supersymmetric gauge theories
\cite{Policastro:2001yc}.

This result is nontrivial because the shear viscosity of a weakly
coupled, perturbative quark-gluon plasma is not small. In fact, the
perturbative result for the shear viscosity, in leading logarithmic
approximation, is \cite{Arnold:2000dr}
\begin{equation}
\eta_C = \frac{d_f T^3}{g^4 \ln g^{-1}} ,
\label{eq:eta-C}
\end{equation}
where $d_f \sim O(100)$ is a numerically determined constant that
weakly depends on the number of quark flavors $n_f$. The result
(\ref{eq:eta-C}), as well as the finding that numerical solutions of
the Boltzmann equation exhibit fluid dynamical behavior only when
the cross section between gluons is artificially increased by a
factor ten or more \cite{Molnar:2001ux}, have invited speculations
that the matter produced at RHIC is a strongly coupled quark-gluon
plasma (sQGP). The possible microscopic structure of such a state
is not well understood at present
\cite{Shuryak:2004tx,Koch:2005vg,Liao:2005pa}.

However, as Berndt M\"uller pointed out in his talk \cite{Asakawa:2006tc}, there exists
an alternative mechanism that may be responsible for
a small viscosity of a weakly coupled, but expanding quark-gluon plasma.
This  mechanism is based on the theory of particle transport in
turbulent plasmas \cite{Dupree:1966,Dupree:1968}. Such plasmas are
characterized by strongly excited random field modes in certain regimes
of instability, which coherently scatter the charged particles and
thus reduce the rate of momentum transport. The scattering by turbulent
fields in electromagnetic plasmas is known to greatly increase the
energy loss of charged particles \cite{Okada:1980} and reduce the
heat conductivity \cite{Malone:1975,Okada:1978} and the viscosity
\cite{Abe:1980a,Abe:1980b} of the plasma. Following Abe and Niu
\cite{Abe:1980b},  the contribution from turbulent fields
to transport coefficients was called ``anomalous''.

The sufficient condition for the spontaneous formation of turbulent,
partially coherent fields is the presence of instabilities in
the gauge field due to the presence of charged particles. This
condition is met in electromagnetic plasmas with an anisotropic
momentum distribution of the charged particles \cite{Weibel:1959},
and it is known to be satisfied in quark-gluon plasmas with an
anisotropic momentum distribution of thermal partons
\cite{Mrowczynski:1988dz,Mrowczynski:1993qm,Romatschke:2003ms}.

The additional contribution to the viscosity, $\eta_B$, induced
by the turbulent fields
decreases with increasing strength of the
fields. Since the amplitude of the turbulent fields grows
with the magnitude of the momentum anisotropy, a large anisotropy
will lead to a small value of $\eta_B$. Because the relaxation rates
due to different processes are additive, the total viscosity is given by
\begin{equation}
\eta^{-1} = \eta_B^{-1} + \eta_C^{-1} .
\label{eq:eta-total}
\end{equation}
This equation implies that $\eta_B$ dominates the total shear viscosity,
if it is smaller than $\eta_C$. In that limit, the anomalous mechanism
exhibits a stable equilibrium in which the viscosity regulates itself:
The anisotropy grows with $\eta$, but an increased anisotropy tends to
suppress $\eta_B$. The result is that in the weak coupling limit, the anomalous viscosity is
much smaller than the viscosity due to collisions among thermal partons.
By reducing the shear viscosity of a weakly coupled, but expanding
quark-gluon plasma, this mechanism could possibly  explain the
observations of the RHIC experiments without the assumption of a
strongly coupled plasma state.

\section{From SQM to F(lavor)QM}

\subsection{Heavy-Quark Production, Diffusion and Energy Loss}

The calculation of heavy quark production in the framework of pQCD
is fairly well established -- recent improvements having 
been reported by I. Vitev at this meeting \cite{Vitev} 
(see also \cite{Vogt:2004hd} and references
therein). Among the novelties presented was the work
by K. Tuchin, who discussed heavy quark production in High Parton Density QCD 
in a quasi-classical approximation, including low-$x$ quantum evolution, 
as well as heavy-quark production
based on the effect of 
pair production in external fields \cite{Tuchin}.

For moderate transverse momenta ($p_T\leq$ a few GeV/$c$) the energy
loss of heavy quarks is thought to be dominated by inelastic collisions
with medium constituents, rather than gluon radiation
\cite{Mustafa:2003vh,Mustafa:2004dr}, because the heavy quarks are not
ultra-relativistic and gluon radiation is suppressed by the so-called
dead-cone effect \cite{Dokshitzer:1991fd}. These findings were
highlighted in a comprehensive analysis of the non-photonic electron
nuclear suppression factors measured at RHIC presented by M. Djordjevic
\cite{Djordjevic,Wicks:2005gt}. The analysis clearly indicates the
importance of including charm and bottom quarks in the calculation and 
taking their radiative as well as collisional energy
loss into account.

The importance of the collisional energy loss contribution
and the possibility of heavy quarks actually thermalizing in the
medium provides an opportunity to utilize them as probes for the transport
coefficients of the QCD medium.
Since collisional energy loss
occurs in many small steps, the motion of a heavy quark can thus be
described by a Fokker-Planck or, equivalently, Langevin equation.
Several studies based on such an approach have recently been done
\cite{Moore:2004tg,vanHees:2004gq,Gossiaux:2004qw}. The most detailed
and extensive one of these was performed by Moore and Teaney \cite{Moore:2004tg},
who derived an expression for the diffusion coefficient $D$ in the
framework of hard-thermal loop (HTL) improved perturbation theory and
discussed the limitations of the Fokker-Planck approach. These authors
also studied the resulting phenomenology of heavy quark transport in
dense matter created by a heavy ion collision, using a boost invariant
hydrodynamical model with an ideal equation of state.
At SQM 2006, heavy quark diffusion calculations were discussed by
P.B. Gossiaux \cite{Gossiaux:2004qw,Gossiaux} as well as by 
R. Rapp \cite{vanHees:2004gq,Rapp:2006ta}. Whereas the philosophy 
of \cite{Gossiaux:2004qw,Gossiaux} is to determine the transport coefficients from a 
comparison to data, the approach of \cite{vanHees:2004gq,Rapp:2006ta} is based
on introducing a resonant charm --  light quark interaction 
and then
calculating the drag- and diffusion coefficients for the Langevin evolution
on the basis of that interaction. A comparison to data would thus yield
information on the microscopic interaction between charm and the light quark
species (see right frame
of figure~2).


\subsection{Charmonium Spectral Functions}
In 1986 Matsui and Satz proposed \cite{Matsui:1986dk} that the suppression of
heavy quarkonia-mesons could provide one of the signatures for
deconfinement in QCD at high temperatures. The idea was based on an analogy
with the well known Mott transition in condensed matter systems.  At high
densities, Debye screening in a quark-gluon plasma
reduces the range of the attractive force
between heavy quarks and antiquarks, and above some critical density
screening prevents the formation of bound states.  The larger bound
states were expected to dissolve before the smaller ones as the
temperature of the system increases.  The $\psi'$ and $\chi_c$ states
were thus expected to become unbound just above $T_c$, while
the smaller $\psi$ state would only dissolve above $\approx 1.2 T_c$.

However, QCD lattice gauge-theory calculations of charmonium correlators
in recent years have necessitated a revision in our understanding of
the dissociation of charmonium states and what it implies for the 
properties of the surrounding medium: as discussed by P. Petreczky at
this meeting, the 1S charmonia states ($J/\psi$, $\psi'$ and $\eta_C$) survive
to unexpectedly high temperatures above 1.5 $T_C$ and only the 
1P states (i.e. the $\chi_C$) dissolves around 1.2 $T_C$ 
\cite{Petreczky,Jakovac:2006bx}. It is therefore questionable whether
the observed charmonium suppression in Pb+Pb collisions 
at the SPS \cite{Abreu:2000ni}
is truly a {\em smoking gun} signature for deconfinement or rather the result
hadronic dissociation (see e.g. \cite{Spieles:1999kp}).
Moving on to RHIC energies, the situation will become even more complicated:
achievable temperatures may be above the threshold for
charmonium dissociation (which is unlikely in the SPS case), however,
the additional suppression may be compensated through novel production 
mechanisms.
The multiplicity of produced charm and anti-charm quarks per event at RHIC
is sufficiently large that charmonium-regeneration via parton recombination
may occur \cite{Thews:2000rj,Grandchamp:2001pf}, giving rise to a fairly flat behavior
of the observed charmonium yield as a function of beam energy.

Another exciting prospect is the measurement of charmonium
elliptic flow, first shown in form of preliminary data by NA60 here
at this conference \cite{Foerster}. The measurement of significant
charmonium elliptic flow would indicate charm thermalization and 
charmonium regeneration via parton recombination. In addition it has
been shown \cite{Rapp:2006ta}, that this observable exhibits a strong sensitivity
to the in-medium interaction of charm quarks.

\section{Outlook}

The field of strangeness in heavy-ion collisions -- and with it this
conference series -- has developed tremendously over the past 20 years.
While strangeness as a QGP signature was instrumental in creating
the SQM conference series, strangeness production
in heavy-ion collisions has become an
impressively versatile tool for the characterization of 
all aspects of confined and deconfined QCD matter,
as witnessed by the work presented
at this conference. The  future of the field will most likely bring the
generalization of  the concepts and lessons learned from strangeness
into the heavy-quark sector at higher incident beam energies. 
The application of flavor-dominated physics phenomena as probes of the
the hot and dense QCD medium will be an exciting field for many
years to come.

\ack  
This work was supported by a DOE Outstanding Junior Investigator Award
under grant number DE-FG02-03ER41239. I wish to thank all 
participants of the SQM 2006 conference for many insightful
discussions without which this write-up would not have
been possible.

\end{document}